\documentclass[sigconf]{acmart}

\newcommand{\skill}[1]{\textcolor{black}{\textsc{#1}}}

\AtBeginDocument{
 \providecommand\BibTeX{{
 \normalfont B\kern-0.5em{\scshape i\kern-0.25em b}\kern-0.8em\TeX}}}

\copyrightyear{2025}
\acmYear{2025}
\setcopyright{cc}
\setcctype{by}
\acmConference[DIS '25]{Designing Interactive Systems Conference}{July 5--9, 2025}{Funchal, Portugal}
\acmBooktitle{Designing Interactive Systems Conference (DIS '25), July 5--9, 2025, Funchal, Portugal}\acmDOI{10.1145/3715336.3735788}
\acmISBN{979-8-4007-1485-6/2025/07}

\usepackage[font=small]{caption}
\usepackage{subcaption}
\usepackage{tabularray}
\usepackage{balance}

\UseTblrLibrary{booktabs}
\usepackage{enumitem}
\newlist{tabenum}{enumerate}{1}
\setlist[tabenum]{wide=0pt,
 nosep, 
 leftmargin= * ,
 label*=\arabic*.,
 after=\vspace{-\baselineskip},
 before=\vspace{-0.6\baselineskip}}

\begin{document}

\title{Rubikon: Intelligent Tutoring for Rubik's Cube Learning Through AR-enabled Physical Task Reconfiguration}

\author{Haocheng Ren}
\authornote{Both authors contributed equally to this research.}
\affiliation{
 \institution{University of Michigan\\
 Computer Science \& Engineering}
 \city{Ann Arbor}
 \state{Michigan}
 \country{USA}
}
\email{chrenx@umich.edu}

\author{Muzhe Wu}
\authornotemark[1]
\affiliation{
 \institution{Carnegie Mellon University\\
 Human-Computer Interaction Institute}
 \city{Pittsburgh}
 \state{Pennsylvania}
 \country{USA}
}
\affiliation{
 \institution{University of Michigan\\
 Computer Science \& Engineering}
 \city{Ann Arbor}
 \state{Michigan}
 \country{USA}
}
\email{muzhew@andrew.cmu.edu}

\author{Gregory Croisdale}
\affiliation{
 \institution{University of Michigan\\
 Computer Science \& Engineering}
 \city{Ann Arbor}
 \state{Michigan}
 \country{USA}
}
\email{gregtc@umich.edu}

\author{Anhong Guo}
\affiliation{
 \institution{University of Michigan\\
 Computer Science \& Engineering}
 \city{Ann Arbor}
 \state{Michigan}
 \country{USA}
}
\email{anhong@umich.edu}

\author{Xu Wang}
\affiliation{
 \institution{University of Michigan\\
 Computer Science \& Engineering}
 \city{Ann Arbor}
 \state{Michigan}
 \country{USA}
}
\email{xwanghci@umich.edu}

\begin{abstract}
Learning to solve a Rubik's Cube requires the learners to repeatedly practice a skill component, e.g., identifying a misplaced square and putting it back. However, for 3D physical tasks such as this, generating sufficient repeated practice opportunities for learners can be challenging, in part because it is difficult for novices to reconfigure the physical object to specific states. We propose Rubikon, an intelligent tutoring system for learning to solve the Rubik's Cube. Rubikon reduces the necessity for repeated manual configurations of the Rubik's Cube without compromising the tactile experience of handling a physical cube. The foundational design of Rubikon is an AR setup, where learners manipulate a physical cube while seeing an AR-rendered cube on a display. Rubikon automatically generates configurations of the Rubik's Cube to target learners' weaknesses and help them exercise diverse knowledge components. In a between-subjects experiment, we showed that Rubikon learners scored 25\% higher on a post-test compared to baselines.
\end{abstract}




\begin{teaserfigure}
\centering
\includegraphics[width=.85\linewidth]{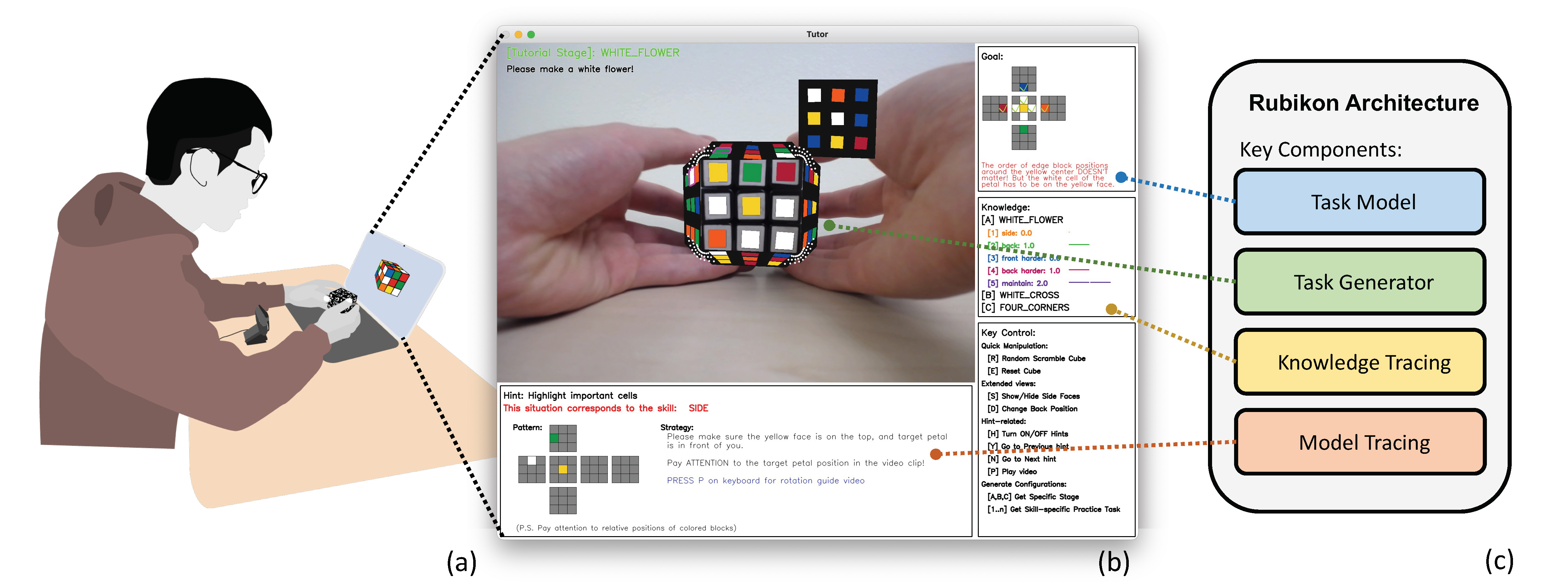}
 \caption{Rubikon is an intelligent tutoring system for Rubik's Cube learning. (a) The foundational design of Rubikon is an AR setup, where learners manipulate a physical cube with ArUco markers attached to each square, and pose a camera towards the cube to enable tracking and rendering. With this setup, learners see a rendered Rubik's Cube on a display while manipulating the physical cube in their hands. (b) Through AR rendering, Rubikon automatically generates new configurations of the Rubik's Cube for the user to practice unmastered skills. Rubikon detects the status of the cube to infer user behavior and provide immediate feedback and hints. (c) Rubikon supports the learning of a 3D physical task by integrating key design principles of cognitive tutors which have seen success in tutoring math and programming.}
 \Description{Teaser diagram labeled with (a), (b), and (c) showing the key components of the Rubikon architecture. Part (a) shows a student learning to solve Rubik's Cube using Rubikon. Part (b) is a zoom-in view of Rubikon's user interface that the student is looking at. Part (c) shows the key components of Rubikon: Task Model, Task Generator, Knowledge Tracing, and Model Tracing. Parts (b) and (c) are connected: Task Model is linked to the goal image on the screen; Task Generator is connected to the Rubik's Cube; Knowledge Tracing is linked to the skillometer panel; Model Tracing is linked to the hint panel.}
 \label{fig:teaser}
\end{teaserfigure}

\maketitle

\section{Introduction}
The Rubik's Cube, a 3D combination puzzle, has been widely recognized for its cognitive and educational benefits \cite{zeng2018overview}, including bolstering people's spatial intelligence \cite{joyner2008adventures}, supporting math \cite{joyner2008adventures}, architecture \cite{bandelow2012inside} and engineering \cite{sen2021computational} learning, and improving people's hand-eye coordination \cite{valerie2020supporting}. Conventional methods of learning to solve a Rubik's Cube involve the use of manuals \cite{rubiksRubiks}, video demonstrations \cite{youtubeSolveRubiks}, and interactive websites \cite{rubikscuRubiksCube}. These methods possess notable limitations. First, it introduces significant cognitive burden on the learners to manually map the status of their cube at hand to the cube in the paper or online tutorials. Second, it lacks repeated configurations of the Rubik's Cube for the learners to practice. Third, given the immense complexity of the Rubik's Cube's configuration space, it is difficult for learners to keep track of their knowledge progression and practice unmastered elements. 

Prior work on cognitive tutors \cite{koedinger1997intelligent, anderson1995cognitive} has demonstrated great success in providing personalized instruction to learners, and has been widely explored in subject domains such as math \cite{olsen2014using, koedinger1997intelligent, koedinger2007exploring, rau2009intelligent} and programming \cite{rivers2017data, price2017isnap}. However, applying the design principles of cognitive tutors in building a tutoring system for the physical Rubik's Cube poses key challenges. A key consideration is to retain the haptic experience in solving the Rubik's Cube while providing the intelligent tutoring affordances through knowledge and model tracing \cite{koedinger1997intelligent, vanlehn2006behavior}. On the one hand, it is essential for cognitive tutors to provide repeated practice opportunities to learners on unmastered knowledge elements. However, in the context of solving a Rubik's Cube, it requires repeated configurations of the cube in order for learners to practice on diverse problem scenarios, which is difficult to achieve, especially for novices. On the other hand, adaptive tutoring requires accurately capturing student performance on the task in order to infer their skill mastery and give feedback. However, it is challenging to acquire learner's step-by-step performance when they are solving a Rubik's Cube. 

In this work, we are driven by two overarching research questions: \textit{(i)} For tasks that require manual configuration of physical objects in order for novices to practice, how might we automatically generate targeted practice opportunities? \textit{(ii)} If we are able to provide these practice opportunities, will they help novices learn? We specifically probed into these research questions in the context of learning to solve the Rubik's Cube, a popular physical task with considerable benefits on spatial thinking skills \cite{zeng2018overview}, hand-eye coordination \cite{valerie2020supporting}, math learning \cite{joyner2008adventures}, and computational thinking \cite{agostinelli2021designing}. Uniquely, Rubik's Cube's symmetrical shape makes it easier to track the positions of each square.\footnote{A Rubik's Cube has six faces, each face has nine squares, also referred to as ``cells.''} This makes the Rubik's Cube a good candidate for developing intelligent tutoring setups, where generating new configurations of the physical object becomes possible and inferring learners' performance becomes easier. 

We present Rubikon, an intelligent tutoring system for learning to solve the Rubik's Cube. The design of Rubikon closely follows the design principles of cognitive tutors \cite{koedinger_cognitive_2006}, while addressing novel technical challenges when implementing them in such a 3D physical task. The foundational design of Rubikon is an AR setup, where learners manipulate a physical cube with ArUco markers attached to each square, and pose a camera towards the cube to enable tracking and rendering. With this setup, learners see a rendered Rubik's Cube on a display while manipulating the physical cube in their hands, as shown in Figure~\ref{fig:teaser}. This enables Rubikon to \textit{(i)} generate new configurations of the Rubik's Cube to present targeted problem-solving opportunities to learners, which eliminates the need to manually reset or reconfigure the cube to a specific state for each new problem; and \textit{(ii)} track the cube's status change to infer users' operations, in order to provide real-time feedback and model their learning trajectory. We implement Rubikon as a full-fledged cognitive tutor: it specifies a task model and compares user behavior against the task model to provide feedback to learners; and it uses a skillometer to track a learner's weaknesses and prompts them to practice more when needed. 

To evaluate whether Rubikon can successfully generate learning opportunities that target learners' weaknesses and thus enhance learning, we conducted a between-subjects experiment with 36 participants who self-identified as novices in solving a Rubik's Cube. The study also aims to investigate whether the AR setup poses extra cognitive load on the users. With these goals, we designed two baseline conditions. Baseline 1 simulates a business-as-usual learning scenario, in which learners watch online tutorial videos while manipulating a regular cube in their hands. Learners can pause at any time and re-watch the video in any way they want. In Baseline 2, learners had the same instructions as Baseline 1, but manipulated a cube augmented with ArUco markers while seeing the rendered cube on a display. The comparison between the two baselines aims to address whether the AR rendering of the Rubik's Cube on a separate display adds to users' cognitive load when their views and hand operations are separated. Finally, in the Rubikon condition, learners learn to solve the Rubik's Cube with the full system functionality, including the AR setup and the intelligent tutoring capabilities, and learners receive feedback on every move and get prompted to repeatedly practice a skill until they demonstrate mastery of it. 

We found that learners in the Rubikon condition scored 25\% higher on the post-test compared to the traditional video tutorial baseline. We further probed into the reasons behind the higher learning gains, and found that learners in the Rubikon condition had more comprehensive and balanced coverage of the diverse skill components required to solve a Rubik's Cube during their practice time, whereas learners in the baselines had limited exposure constrained by the physical configurations of the cube. Many participants in the baselines struggled to match the configuration of the cube in their hands with that in the tutorial video, which increased the ``preparation cost'' for learners to engage in meaningful practice. Additionally, learners in the three conditions reported a similar level of cognitive load. We demonstrate the feasibility and pedagogical effectiveness of automatically generating targeted and personalized practice opportunities for Rubik's Cube learning. 

With the acknowledgment that the AR-setup and user behavior tracking techniques implemented in Rubikon are specifically limited to the task of learning to solve a Rubik's Cube, the concept of generating targeted and personalized practice opportunities for physical task learning while retaining the haptic experience through AR may apply in other contexts. Through this proof-of-concept study, we encourage future Mixed Reality-based tutoring systems for physical task learning to engage with the idea of personalized task reconfiguration. 

\section{Related Work}

\subsection{Rubik's Cube Learning}\label{sec:rubikscubelearning}
In his 1981 book ``Magic Cube'' \cite{singmaster1981notes}, Singmaster proposed the layer-by-layer strategy, which involves solving the Rubik's Cube starting with the first layer and then progressing to the second and third. This method has become the preferred strategy for beginners to learn. For a long time, paper manuals (e.g., \cite{rubiksRubiks}), typically included with a purchased Rubik's Cube, have been the primary learning resource. As technology advanced, video demonstrations (e.g., \cite{youtubeSolveRubiks}) and interactive websites (e.g., \cite{learnhowtosolvearubikscubeSolveRubiks, rubikscuRubiksCube} offering virtual simulations of the cube and interactive tutorials) became increasingly popular. However, these methods require learners to constantly match their Rubik's Cube with the Rubik's Cube in the tutorials, creating extraneous cognitive load.

There are also existing commercial and research solutions for tracking the cube's state using internal \cite{giiker_supercube_i3s,particula_gocube} and external sensors \cite{park2016augmented,buildits_2018, hale2022diy}. Products like GIIKER SUPER CUBE~\cite{giiker_supercube_i3s} and GoCube~\cite{particula_gocube} enable tracking through embedded Bluetooth or IMU sensors to provide real-time guidance through connected mobile devices. Besides tracking (e.g., with RGB cameras~\cite{park2016augmented,jokinen2024solving})), prior work in academic research has also focused on making Rubik's Cube learning more accessible and engaging. Khan et al. elicited design guidelines for enhancing learners' engagement with the Rubik's Cube through a series of workshops with designers and educators~\cite{khan2023participatory}. Chen and Liu proposed a mathematical method for learning path optimization that balances knowledge density with playfulness~\cite{chen2019analysis}. However, all the prior solutions still require the users to manually reset or reconfigure the cube to specific states in order to exercise certain knowledge components, which is particularly difficult for novices. Rubikon's AR setup using ArUco markers enables \textit{digital reconfiguration} of the cube state without requiring any \textit{physical manipulations}, which is complementary to the existing approaches for state tracking, and allowing for more flexible implementations of intelligent tutoring systems to target learner weaknesses by generating any targeted exercises on-the-fly.

\subsection{Cognitive Tutor Design Principles}
Decades of research has shown that cognitive tutors \cite{koedinger1997intelligent, koedinger_cognitive_2006, pane2014effectiveness} lead to better learning in comparison with traditional methods. We briefly introduce the design principles of cognitive tutors as summarized in prior work \cite{koedinger_cognitive_2006}. First, cognitive tutors represent student competence as a set of production rules. The production rules are often derived from cognitive task analysis \cite{kacprzyk_rule-based_2010} which specifies the knowledge components one needs to acquire in order to complete a task successfully. Second, cognitive tutors provide instruction in a problem-solving context \cite{koedinger1997intelligent}. Students need to apply the production rules in order to solve the problems. Students acquire the production rules, i.e., the skills, through repeatedly applying them during problem solving \cite{koedinger1997intelligent, koedinger1998illustrating}. Third, model tracing and knowledge tracing \cite{koedinger_cognitive_2006} are two main algorithms used in cognitive tutors. Model tracing compares the students' move with the projected move based on an expert task model to give them hints and feedback \cite{koedinger1997intelligent, koedinger_cognitive_2006}. Knowledge tracing evaluates students' mastery of each skill based on a sequence of student performance \cite{pavlik2009performance, koedinger_cognitive_2006}. Rubikon carefully follows the above design principles in defining a task model, generating new problem-solving scenarios through AR-enabled automatic reconfiguration, and providing knowledge and model tracing. 

Nevertheless, the implementation of cognitive tutor design principles on a 3D physical task (such as solving Rubik's Cube) poses several challenges. First, providing learners with problem-solving tasks in an authentic context can be challenging when the physical objects require repeated configuration. Second, both model tracing and knowledge tracing require capturing student moves, i.e., their performance on a specific task. In prior tutors, e.g., algebra \cite{pane2014effectiveness}, programming \cite{price2017isnap}, and chemistry \cite{king2022open} tutors, learners can demonstrate their performance through web-based \cite{rau2009intelligent, rivers2017data} or conversational interfaces \cite{graesser2005autotutor}. However, for 3D physical tasks, how to capture learners' performance accurately becomes a challenge.

\subsection{Mixed Reality-based Physical Task Guidance and Tutoring Systems}
People traditionally turn to text- and video-based tutorials for guidance on physical tasks \cite{chi2012mixt, yamaguchi_video-annotated_2020}. As an emergent technology, Mixed Reality (MR) shows prospects of revolutionizing the physical task learning experience, allowing students to perceive virtual instructions while interacting with the tangible world, leading to increased engagement \cite{parong2018learning} and improved learning outcomes \cite{radu2014augmented, funk2016interactive, tang_comparative_2003}. Additionally, prior work has shown the importance of immersion for physical task learning in which learners are required to physically coordinate their body and hands \cite{patel2006effects}. Here, we review prior MR-based physical task guidance and tutoring systems on three aspects: \textit{(i)} how physical tasks are reconfigured, \textit{(ii)} how learners' performance is captured, and \textit{(iii)} what feedback is provided to learners. 

Many prior physical task performance support systems are built around linear and single-path tasks, e.g., machine operation \cite{liu2023instrumentar, huang2021adaptutar, cao_mobiletutar_2022, chidambaram2021processar, kong2021tutoriallens} and assembly \cite{yamaguchi_video-annotated_2020, whitlock_authar_2020}. These tutorial systems do not have the affordance for reconfiguration. Other systems focus on non-linear tasks with divergent procedures, including learning musical instruments \cite{thoravi2019loki}, physical exercises \cite{faridan2023chameleoncontrol, monteiro2023teachable, anderson2013youmove}, and combination games \cite{gupta_duplotrack_2012}. However, these systems often do not generate problem-solving scenarios that pertain to the learners' prior knowledge. They either require the learners to select tasks themselves or demand human tutors' real-time, one-on-one instruction which is not always accessible \cite{faridan2023chameleoncontrol, thoravi2019loki}. Thus, automatically generating configurations of physical objects to enable deliberate practice is relatively under-explored in MR-based guidance and tutoring systems. 

Most existing MR physical task tutors employ sensor-based methods \cite{liu2023instrumentar}, computer vision models \cite{huang2021adaptutar, monteiro2023teachable}, and other advanced motion-trackers \cite{faridan2023chameleoncontrol, thoravi2019loki} to monitor users' behavior, which often demands expensive setup and calibration efforts, and is prone to errors. Rubikon explores a lightweight and accurate tracking method inferring users' behavior through ArUco markers attached to each square of a Rubik's Cube, which also enables task reconfiguration on-the-fly.

Prior work has extensively examined the effectiveness of different types of feedback in MR-based guidance and tutoring systems, including visual cues (e.g., words \cite{cao_mobiletutar_2022, whitlock_authar_2020, rajaram2022paper}, 2D figures and videos \cite{chidambaram2021processar, cao_mobiletutar_2022, anderson2013youmove, thees2020effects}, 3D arrows \cite{chidambaram2021processar, cao_exploratory_2020}, and avatars \cite{cao_exploratory_2020, eckhoff2018tutar}) for motion guidance. Recent work Paper Trail \cite{rajaram2022paper} allows teachers to author explanations in an AR environment.
However, most of the referenced systems offer such feedback as static information prior to students' actions and are not responsive. A small fraction of prior work provides interactive feedback \cite{rigby2020piarno, gupta_duplotrack_2012, monteiro2023teachable, liu2023instrumentar}. The most common format is corrective feedback (e.g., warning signs \cite{liu2023instrumentar}) and on-demand hints (e.g., animations \cite{huang2021adaptutar}). Through model tracing, Rubikon provides interactive feedback and multi-level hints to users in support of their learning. 

\begin{figure*}[t!]
\includegraphics[width=0.8\linewidth]{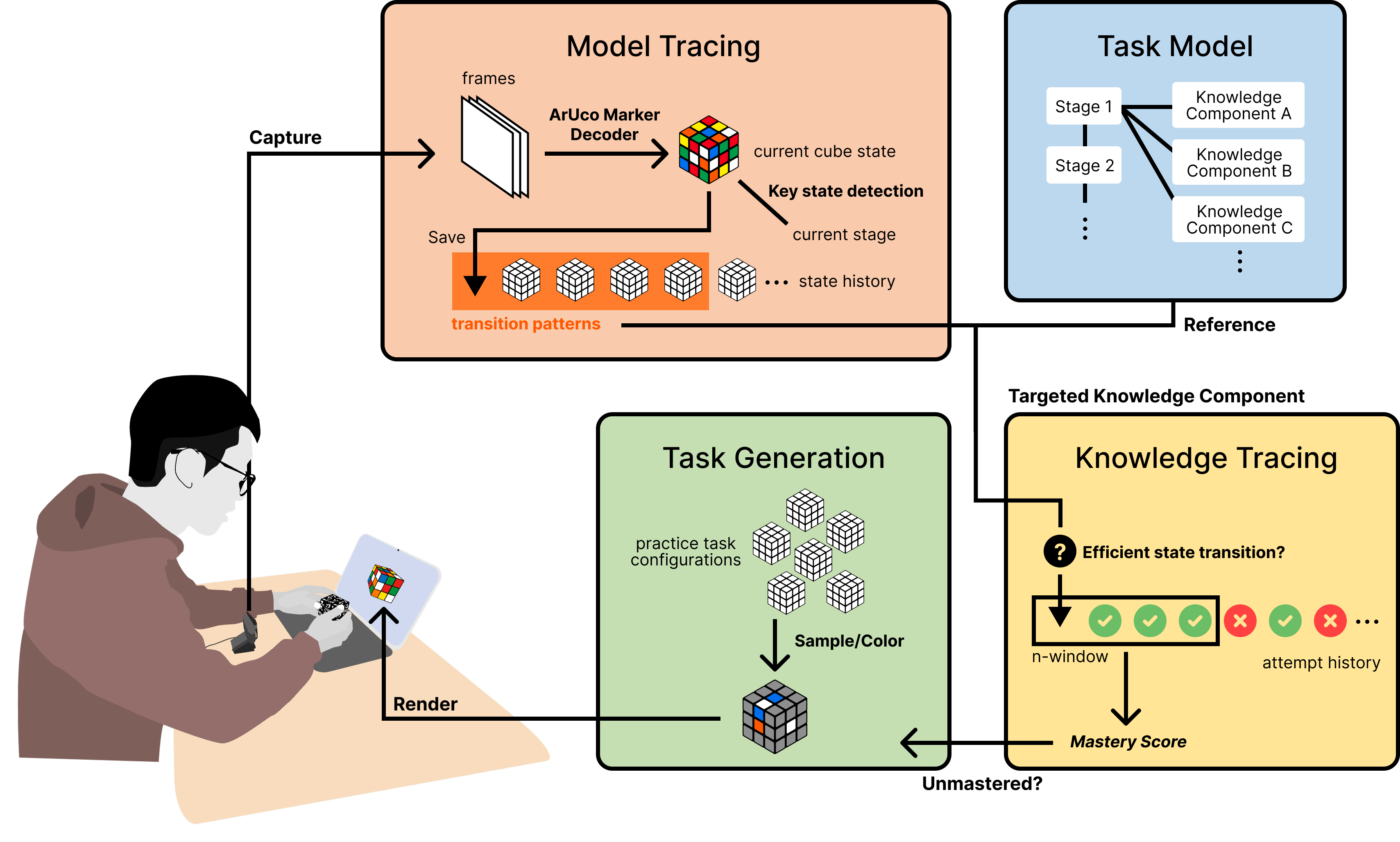}
\vspace{-1pc}
 \caption{Rubikon enables personalized learning through a closed-loop design. While the learner manipulates the cube, the Model Tracing module parses frames streamed from a webcam into cube states. Informed by a Task Model that defines the movement sequences learners need to perform, the Knowledge Tracing module analyzes the transitions between cube states to assess mastery. For knowledge components not yet mastered, the Task Generation module correspondingly samples cube configurations and renders them on the screen for deliberate practice.
 }
 \Description{A diagram showing a learner manipulating a physical Rubik's Cube in front of a laptop. The feed of a webcam is processed by a Model Tracing module to identify cube states. These states are analyzed by a Knowledge Tracing module, which assesses mastery based on cube transitions. A Task Generation module uses this assessment to display new cube configurations on a screen for the learner to practice.
 }
 \label{fig:rubikon-arch}
\end{figure*}

\section{Rubikon: An Intelligent Tutoring System for Rubik's Cube Learning}
We introduce Rubikon, an intelligent tutoring system for learning to solve the Rubik's Cube, that is built using Python and OpenCV. The design of Rubikon closely follows the design principles of cognitive tutors \cite{koedinger_cognitive_2006}, while addressing novel technical challenges when implementing such principles in a 3D physical task to learn to solve a Rubik's Cube. Note, here are the often-used terminologies of a Rubik's Cube: \textit{(i)} square: a Rubik's Cube has six faces, and each face has nine squares; \textit{(ii)} block: a sub-cube of the Rubik's Cube. A Rubik's Cube has 27 blocks. A block can have three squares when it is at the corner, two squares when it is in the middle, or one square when it is in the center of a face; and \textit{(iii)} petal: it refers to the blocks with two squares that connect two faces of the cube.

\subsection{AR Rendering in Support of Task Generation}
\label{sec:rendering}
It is uniquely challenging for tutoring systems of 3D physical tasks to provide deliberate practice opportunities as it requires substantial efforts to reconfigure the physical objects to a certain state. In the case of solving a Rubik's Cube, it would be overwhelmingly difficult for novices to reset the cube to some stage due to their lack of skills. 

Rubikon enables automatic task generation through AR rendering. ArUco markers are attached to each square of a Rubik's Cube. Users can pose a camera towards the cube, and the camera detects the markers and renders the color of each square in real time. The learner sees a fully-rendered Rubik's Cube through a display, while manipulating the physical ArUco marker cube in their hands. This setup enables Rubikon to automatically configure the status of a Rubik's Cube and generate tasks for users to solve. A fully rendered Rubik's Cube is shown in Figure \ref{fig:cubes}.

With this setup, Rubikon generates tasks that target the skills the user has not demonstrably mastered. For example, if the user has not mastered solving petals in the \skill{Back Harder} position (see Figure \ref{fig:task_model}), Rubikon generates Rubik's Cube configurations that put a petal in that position for the learner to practice.

\subsection{Capturing Learner Performance to Enable Intelligent Tutoring}
\label{sec:model-tracing}
It is critical for intelligent tutoring systems to track and analyze students' problem-solving processes to offer personalized feedback, and dynamically present new problems to students based on their prior knowledge. Implementing model tracing and knowledge tracing on 3D physical task tutoring is challenging since there are no given ways to capture learner performance on a task. Rubikon uses the status of the Rubik's Cube to infer the learner's behavior, which is convenient and accurate. With the AR setup, the status of the physical cube is accurately tracked with the ArUco markers, making it possible to capture learner behavior and task performance. 

\subsection{Key Components in Rubikon}
With the AR setup as the foundation, we develop Rubikon following the design principles of cognitive tutors \cite{koedinger_cognitive_2006}. Rubikon has the following key components, as shown in Figure~\ref{fig:rubikon-arch}:

\begin{enumerate}
 \item \textbf{Task Model}: It specifies the solution paths to reach a desired state. 
 \item \textbf{Model Tracing}: Rubikon captures users' rotations through detecting the changes of the cube's status. This allows Rubikon to provide real-time feedback on a user's moves by comparing them against the pre-defined task model. 
 \item \textbf{Knowledge Tracing}: Rubikon uses a skillometer to track a user's weaknesses and prompts them to practice more when needed. 
 \item \textbf{Task Generation}: Rubikon provides learners with repeated and personalized practice opportunities by generating new configurations of the Rubik's Cube through AR rendering. 
\end{enumerate}

\subsubsection{\textbf{Task Model}}
\label{learningObjectives}
\begin{figure}[b!]
\includegraphics[width=0.75\linewidth, keepaspectratio]{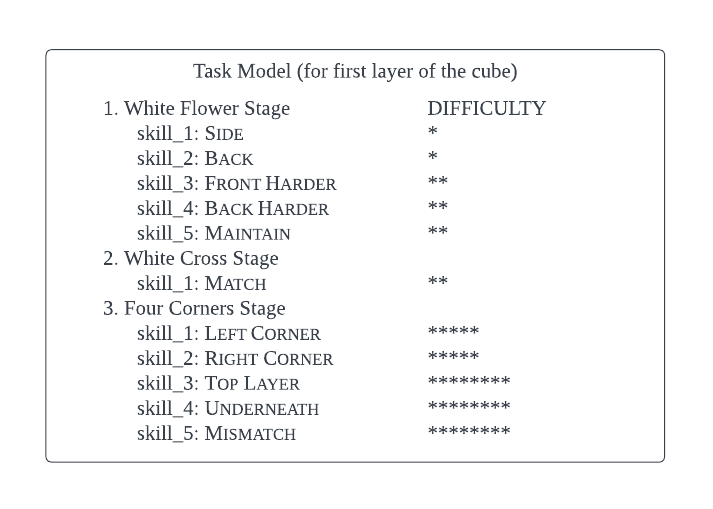}
 \caption{The Rubikon task model outlines 11 essential components for solving the first layer of the Rubik's Cube. Each component corresponds to a specific move sequence necessary to master distinct patterns. Difficulty levels are indicated by the number of *.}
 \Description{List of tutorial stages and their corresponding skills in the Task Model. Three stages include White Flower, White Cross, and Four Corners stages in order. White Flower stage has five skills: SIDE, BACK, FRONT HARDER, BACK HARDER, and MAINTAIN. White Cross stage has one skill called MATCH. Four Corners stage has five skills: LEFT CORNER, RIGHT CORNER, TOP LAYER, UNDERNEATH, and MISMATCH. On the right side of each skill are the stars indicating its difficulty level. Under the same stage, the skills are put in ascending order of difficulty level.}
 \label{fig:task_model}
\end{figure}

The task model is core to any given cognitive tutor, which specifies the skill components and solution paths needed to successfully perform the task. In designing cognitive tutors, people perform a cognitive task analysis of a task domain in order to specify the key components one needs to master, which are referred to as knowledge components \cite{koedinger2012knowledge}. In designing Rubikon, we performed a theoretical cognitive task analysis \cite{kacprzyk_rule-based_2010, lovett1998cognitive} based on the 8-stage conventional solution \cite{learnhowtosolvearubikscubeSolveRubiks} of a Rubik's Cube. We outline 11 knowledge components for solving the first layer of a Rubik's Cube (as shown in Figure \ref{fig:task_model}) and split them into three sequential stages, White Flower, White Cross, and Four Corners. Each knowledge component defines a sequence of moves a user needs to apply and master in order to solve a unique pattern of the cube. It is worth noting that not all knowledge components listed in Figure \ref{fig:task_model} may be encountered by a learner during practice. Depending on the initial configuration of the Rubik's Cube, learners may only encounter the easiest knowledge component, such as \skill{Side} in the first stage. This introduces limitations that even when learners can successfully complete a White Flower, they may not have fully exercised all the knowledge components in this stage. 

\begin{figure}[t!]
\includegraphics[width=0.8\linewidth]{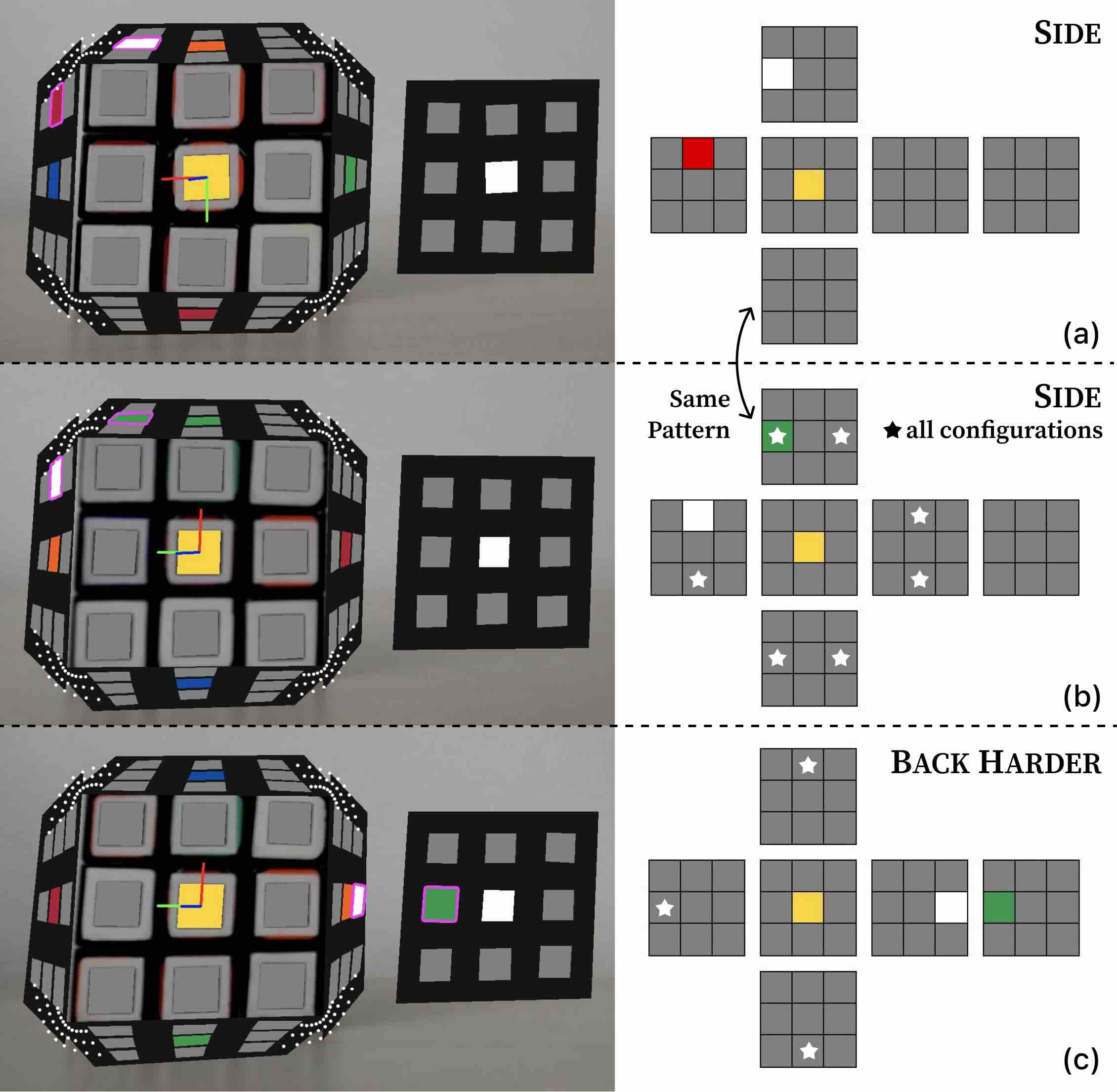}
 \caption{Each knowledge component in Rubikon's task model has a unique pattern of configuration on the Cube. (a) show a specific configuration of the Rubik's Cube corresponding to the knowledge component \skill{Side}. (b) shows that there are 8 configurations that follow this pattern, which all enable the learners to exercise the knowledge component \skill{Side} (indicated by the white stars). (c) shows 4 configurations of the knowledge component \skill{Back Harder} (indicated by the white stars). 
 }
 \Description{Images showing the user interfaces associated with specific skills. The upper images of (a), (b), and (c) show the rendering of the Rubik's Cube through augmented reality with extended views. The bottom images of (a), (b), and (c) show the 2-dimensional cube maps.}
 \label{fig:task_model_skill}
\end{figure}

\textbf{Knowledge components.} \label{kC_details}
Each knowledge component corresponds to a unique pattern of configuration on the Rubik's Cube, and learners need to know what sequence of moves to apply in each situation. There are 11 knowledge components representing over 700K possible configurations for the first layer of the cube since many configurations share the same pattern~\cite{joyner1996mathematics}. For instance, as shown in Figure~\ref{fig:task_model_skill}, for the knowledge component \skill{Side}, although cubes in (a) and (b) are two different configurations, they follow the same pattern. There are 8 possible configurations (indicated by white stars and the white square position in (b)'s pattern) that would allow a learner to exercise the knowledge component \skill{Side}. They share the pattern that the target white square is located in the center on adjacent sides of the yellow-center face. Similarly, (c)'s pattern shows 4 configurations that would allow a learner to exercise the knowledge component \skill{Back Harder}. They share the pattern that the target white square is located at the same horizontal or vertical level as the yellow square, on adjacent sides of the yellow-center face, and has 3 other-colored squares between the white and yellow squares. These patterns essentially define the knowledge components and serve as the basis for Rubikon to generate new configurations that can help learners practice. 

\textbf{Difficulty level of knowledge components.} 
The difficulty levels of the knowledge components are determined by the number of times a block needs to be switched from one layer to a different layer during rotations. For example, in Figure \ref{fig:task_model_skill}, (a) represents the knowledge component \skill{Side}, which has a difficulty level of 1 star (the easiest) because learners only need to rotate the top layer counterclockwise to align the target squares. On the other hand, (c) represents the knowledge component \skill{Back Harder}, which requires users to rotate the rightmost layer clockwise and then the top layer clockwise, making it harder.

\subsubsection{\textbf{Model Tracing}}
When a learner uses Rubikon, the model tracing algorithm detects and tracks the status of the cube through the ArUco markers. The algorithm then infers the user movement (e.g., left-face 90-degree clockwise rotation) based on the cube status change and the rotation templates. 
The system stores the states of the cube after every user move, which we refer to as cube state history. When a block is detected to be put in place, Rubikon gives positive audio feedback to users. When users are struggling, they can request hints. Rubikon provides multi-level hints to the user as in traditional cognitive tutors. 

\subsubsection{\textbf{Knowledge Tracing}}
\label{sec:knowledge-tracing}
Rubikon detects the knowledge component the user just exercised for the purpose of knowledge tracing. The knowledge tracing algorithm analyzes the cube state history and identifies sequences of states where the initial and ending states match those of a specific knowledge component, as specified in the task model. The identified sequence of states signifies an attempt $A$ at a knowledge component. To ``grade'' whether this attempt is correct, the knowledge tracing algorithm examines the sequence of states the Rubik's Cube goes through during this attempt: it first computes the minimum number of steps $\textsc{MinSteps}(\cdot)$ it takes to transition each state $S_i$ to the final state $S_k$; it then calculates the ratio of adjacent states ($S_{i}$, $S_{i+1}$) where the pre-calcuated minimum number of steps is decreasing. If and only if the ratio is higher than a threshold $t_1 = 0.8$, the algorithm considers this attempt at the target knowledge component to be successful. With this, the algorithm can effectively identify situations where the learner, lacking a mental model of which moves to apply, randomly manipulates the cube and accidentally puts a block back in place.

\begin{gather}
 Ratio = \frac{\sum_{i=1}^{k-1}\llbracket \textsc{MinSteps}(S_i,S_k)< \textsc{MinSteps}(S_{i+1},S_k)\rrbracket}{k}, \\ 
 A = \llbracket Ratio > t_1 \rrbracket.
\end{gather}

To model the learner's mastery of each knowledge component, the knowledge tracing algorithm aggregates the most recent n (set at 3) attempts $A\{n\}$, each multiplied by a weight $w(H_i)$ determined by the hint requested $H_i$, if any, for that attempt. Different hint levels correspond to different weights ($w(\text{None})\!=\!1.0;\ w(1)\!=\!0.8;\ w(2)\!=\!0.5;\ w(3)\!=\!0$). The learner is considered to have mastered the knowledge component if and only if the score for that knowledge component exceeds the threshold $t_2$ (set at 2.4).

\begin{align}
 Mastery = \llbracket Score > t_2\rrbracket = \llbracket \sum_{i=1}^n w(H_i)\cdot A_i > t_2\rrbracket.
\end{align}

\begin{figure*}[t!]
\centering
\includegraphics[width=1\linewidth]{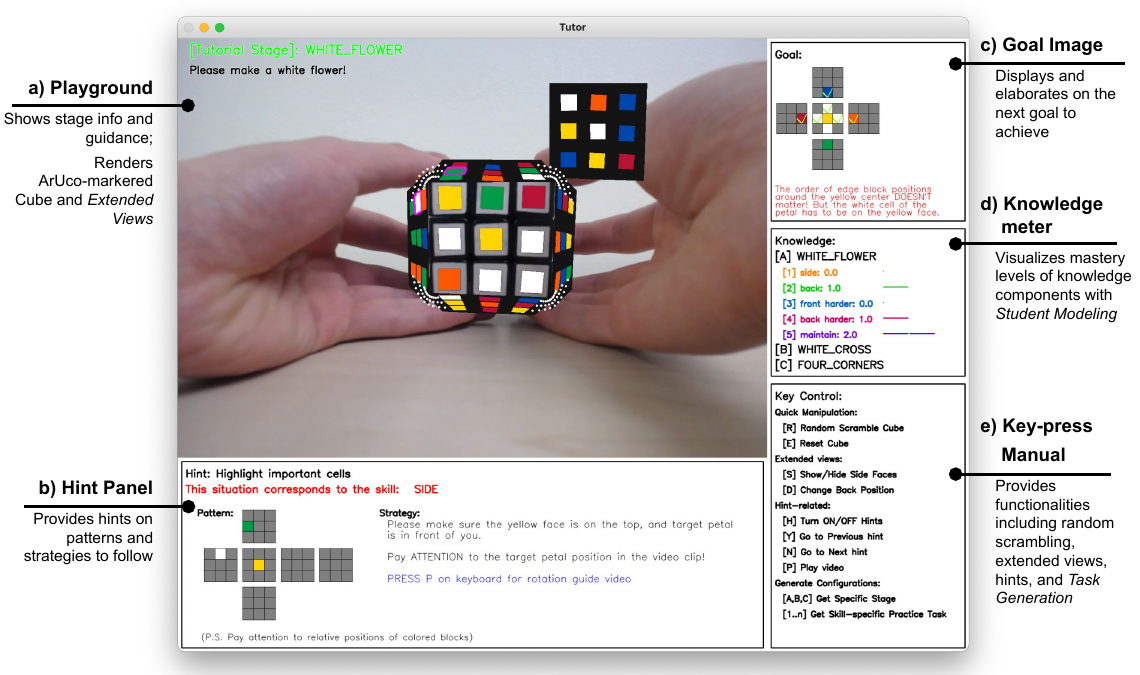}
\caption{The Rubikon user interface has 5 key sections.}
 \label{fig:interface}
 \Description{User interface of Rubikon with five parts labeled (a), (b), (c), (d), and (e). Part (a), Playground, displays stage info and text guidance and renders the ArUco-markered cube with extended views. Part (b), Hint Panel, explains hints upon request, including the pattern and strategy. Part (c), Goal Image, displays and elaborates on the goal that students want to achieve at each stage. Part (d), Skillometer, visualizes students' mastery of knowledge components based on knowledge tracing. Part (e), Controls, facilitates students' needs, including random scrambling, extended views, hint requests, and task generation.}
\end{figure*}

The user's progress is visualized through a skillometer as shown in Figure~\ref{fig:interface}, and all parameters ($n$, $w(\cdot)$, $t_1$, $t_2$) are empirically determined through pilot studies. It is worth noting that knowledge tracing refers to a family of methods on observing, representing and quantifying a student's knowledge states \cite{corbett1994knowledge, abdelrahman2023knowledge}, i.e., the student's mastery levels on the knowledge and skills to be taught. We selected a straightforward method to infer the student's mastery level on each knowledge component, i.e., the student needs to demonstrate several ``good'' sequences of moves. Through our pilot studies, we found it worked well and supported our goals. Future work could use more sophisticated methods as summarized in this recent survey \cite{abdelrahman2023knowledge} on knowledge tracing to more accurately model a student's knowledge states. 

\subsubsection{\textbf{Task Generation}}
With knowledge tracing, the system keeps track of the skills the learner has not mastered and generates new configurations of the Rubik's Cube to target the learner's weaknesses. For example, if the learner has not demonstrably mastered the skill of \skill{Back Harder}, Rubikon will generate configurations of the Rubik's Cube that match the pattern of \skill{Back Harder} for the learner to practice.
These configurations are randomly sampled from a set of cube states associated with each knowledge component, with the specific color arrangements also randomized at runtime, ensuring task variety.

\subsection{Rubikon User Interface}
\label{sec:UI}
The Rubikon user interface contains 5 sections, as shown in Figure~\ref{fig:interface}:

\begin{enumerate}
 \item \textbf{The Playground.} The top-left section displays the rendered Rubik's Cube with extended views. 

 \item \textbf{Hint Panel.} The bottom-left section describes the hints, i.e., what the user should do to achieve the goal. 

 \item \textbf{Goal Image.} The top-right section provides a 2D view of the desired cube state to achieve in the current stage. Only the relevant squares are shown, while the rest of the squares are grayed out.

 \item \textbf{Skillometer.} The middle-right section visualizes the skillometer based on the knowledge tracing algorithm. 

 \item \textbf{Controls.} The bottom-right section lists the key-press events users can trigger. This includes resetting and scrambling the cube, turning on and off extended views, requesting hints, and generating tasks. 
\end{enumerate}

\subsubsection{Extended Views}
\label{sec:extended}
\begin{figure}[b!]
\minipage{\columnwidth}
 \minipage{0.3\columnwidth}
 \includegraphics[width=\columnwidth]{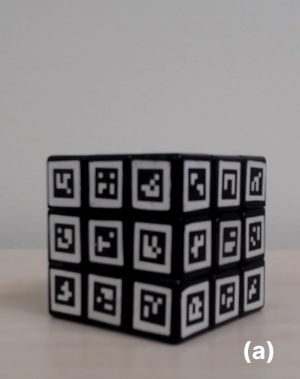}
 \endminipage
 \hfill
 \minipage{0.3\columnwidth}
 \includegraphics[width=\columnwidth]{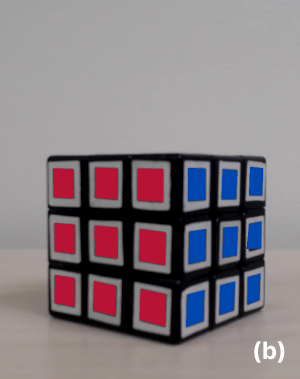}
 \endminipage
 \hfill
 \minipage{0.3\columnwidth}
 \includegraphics[width=\columnwidth]{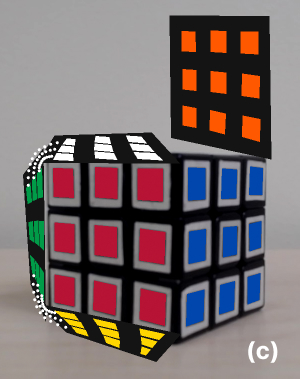}
 \endminipage
\endminipage
 \caption{(a) A physical Rubik's Cube with ArUco markers attached to all squares. (b) A fully-rendered Rubik's Cube on the display with AR rendering. (c) A fully-rendered Rubik's Cube with extended views.}
 \Description{Pictures of 3 different Rubik's Cubes. On the left, it is Rubik's Cube with squares covered by ArUco markers. In the middle, the Rubik's Cube is attached with color using augmented reality rendering. On the right, the Rubik's Cube is further attached with four extended views, including the upper, left, bottom, and back faces.}
 \label{fig:cubes}
\end{figure}

Rubikon provides extended views of the cube to reduce the efforts on frequent rotations. These views, along with AR rendering, provide a comprehensive multi-angular perspective, revealing hidden sides. Through iterative design, we crafted the extended views: 4 trapezoidal shapes for the left, right, up, and down sides, and a shrunk mirrored image for the back side (Figure~\ref{fig:cubes}c). While the default back view location is top-right, users can relocate it for convenience. Spatial connectivity between squares of the same block in the extended views is denoted by arc-shaped dotted lines in Figure~\ref{fig:hint_level}. These views are dynamically computed and adapted to various cube orientations.

\subsubsection{Multi-level Hints}
\label{sec:hints}
\begin{figure}[t!]
\minipage{0.7\columnwidth}
 \centering
 \begin{subfigure}[t]{0.49\columnwidth}
 \centering
 \includegraphics[width=\textwidth]{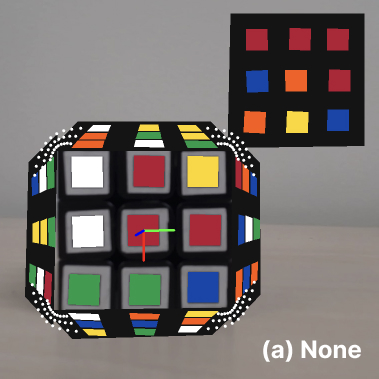}
 
 \end{subfigure}
 \hfill
 \begin{subfigure}[t]{0.49\columnwidth}
 \centering
 \includegraphics[width=\textwidth]{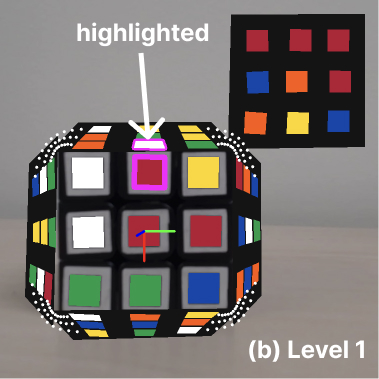}
 
 \end{subfigure}
 
 \begin{subfigure}[t]{0.49\columnwidth}
 \centering
 \includegraphics[width=\textwidth]{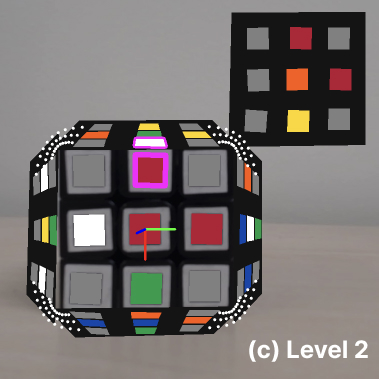}
 
 \end{subfigure}
 \hfill
 \begin{subfigure}[t]{0.49\columnwidth}
 \centering
 \includegraphics[width=\textwidth]{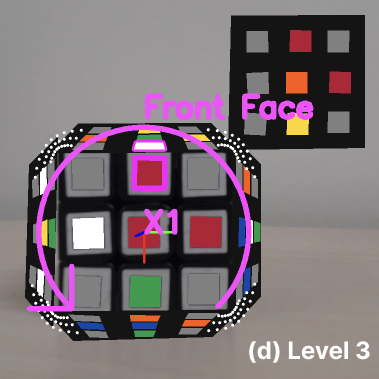}

 \end{subfigure}
 \endminipage

 \caption{Three levels of hint messages: (a) No hint. (b) Hint Level 1: target block is highlighted. (c) Hint Level 2: irrelevant blocks are grayed out. (d) Hint Level 3, step-by-step arrow guidance.}
 \Description{Pictures of Rubik's Cubes (a), (b), (c), and (d) with hint level indicated at the bottom right corner of each picture. (a) has no hint turned on. The color of the Rubik's Cube is rendered in a normal way. (b) has the hint level 1. Two squares are highlighted in this picture. (c) has the hint level 2. Two squares are highlighted, and some squares are grayed out in this picture. (d) has the hint level 3. Two squares are highlighted, some squares are grayed out, and there is a curved arrow with text saying ``front face'' and text indicating ``times one'' next to it.}
 \label{fig:hint_level}
\end{figure}

Rubikon provides three levels of hint messages (Figure~\ref{fig:hint_level}):

\begin{enumerate}
 \item \textbf{Level 1:} Highlight the target block and destination.
 \item \textbf{Level 2:} Gray out unimportant blocks.
 \item \textbf{Level 3:} Explicitly show step-by-step arrow guidance (bottom-up hints).
\end{enumerate}

The information contained in each level of hints increases based on the design principles of cognitive tutors \cite{koedinger_cognitive_2006}. Users have full control over when to request a hint, and at which level. Requesting a hint will lower the user's score on the corresponding knowledge component (refer to Section \ref{sec:knowledge-tracing} for details) indicating the need for further practice. 

\subsubsection{Exploration and Practice Modes}
Users will by default start using Rubikon in the exploration mode, guided by a goal image of the current stage. They can then freely rotate the Rubik's Cube to achieve that goal. Throughout the process, users may request hints and get feedback. After a full exploration, users are directed to the practice mode, during which they are given additional practice opportunities on the knowledge components that they have not demonstrably mastered. Users can also manually request to start the practice mode at any time. 

\section{Evaluation Study}
The evaluation study aims to investigate whether the AR setup in Rubikon AR can successfully generate learning opportunities that target learners' weaknesses and enhance learning. We compare Rubikon with two baselines and address the following research questions:
\begin{itemize}
 \item RQ1: When users learn to solve a Rubik's Cube with the three methods, business-as-usual video tutorials with a regular Rubik's Cube, video tutorials with a rendered Rubik's Cube, and the Rubikon system, which leads to higher learning gains and why?
 \item RQ2: Whether the AR setup poses extra cognitive load on the users?
 \item RQ3: What are users' experiences when using Rubikon? What challenges do they encounter and what are the design implications for developing intelligent tutoring systems for learning 3D physical tasks?
\end{itemize}

\subsection{Study Design}
\begin{figure}[b!]
 \centering
 \begin{subfigure}[t]{0.32\linewidth}
 \includegraphics[width=\textwidth, keepaspectratio]{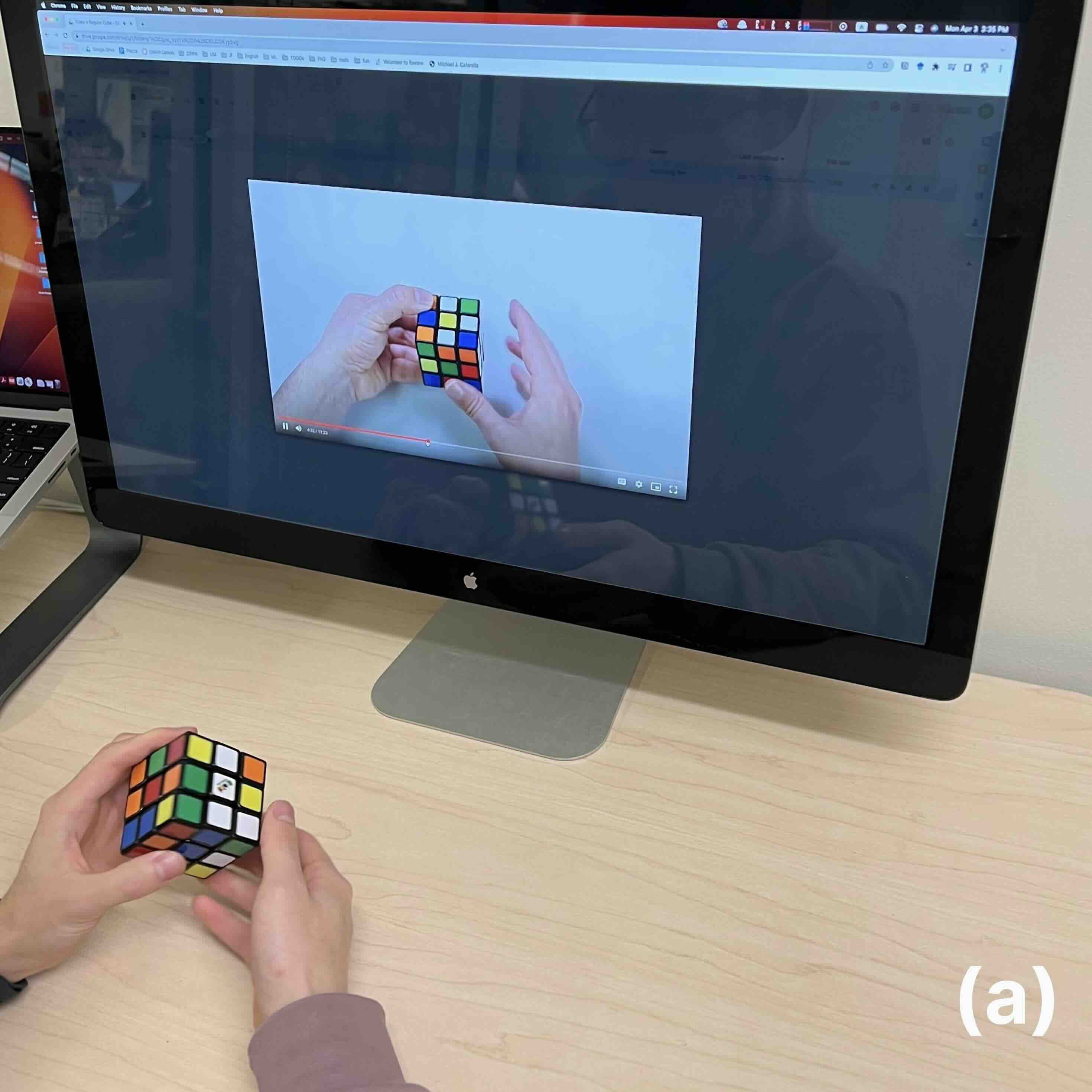}

 \label{fig:image1}
 \end{subfigure}
 \hfill
 \begin{subfigure}[t]{0.32\linewidth}
 \includegraphics[width=\textwidth]{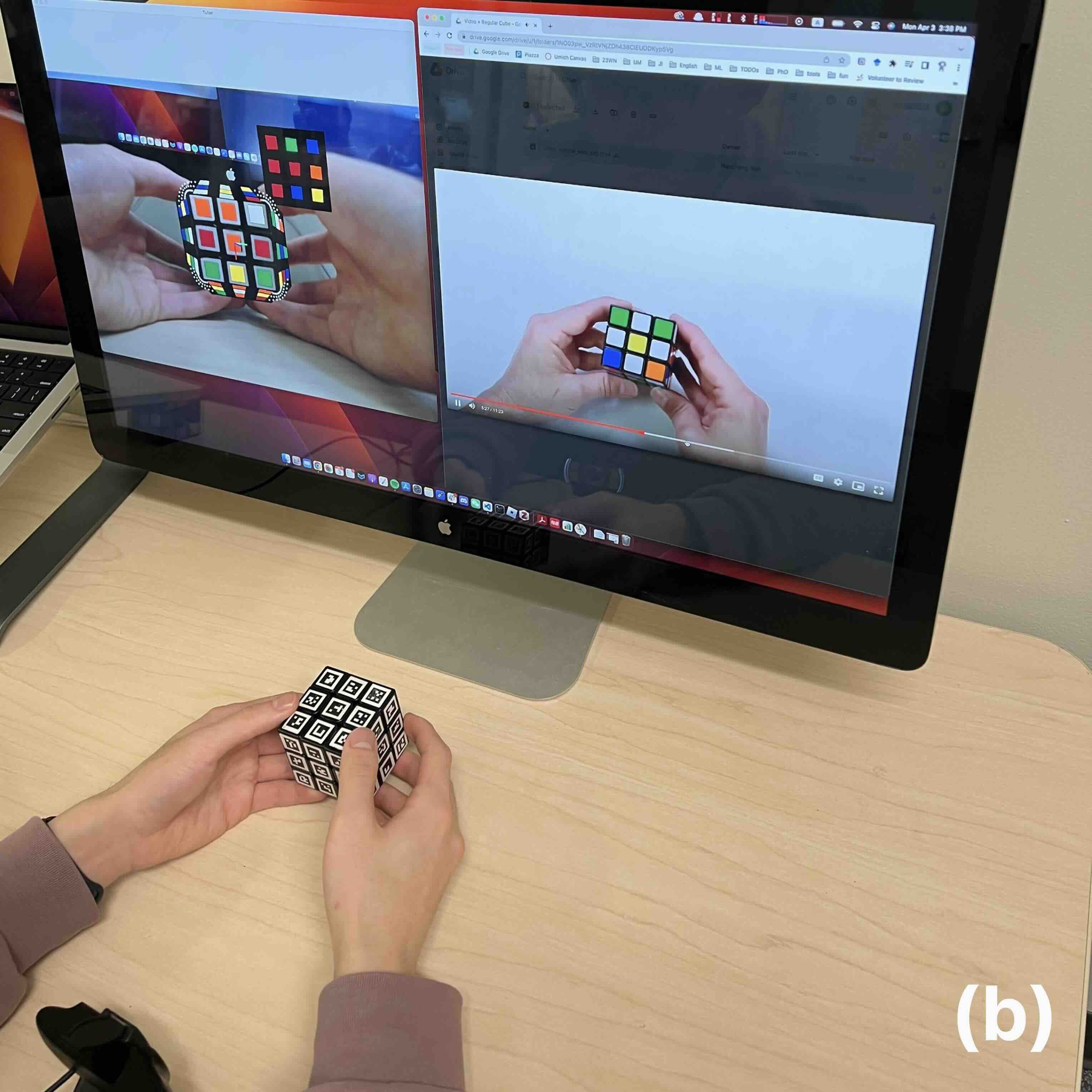}

 \label{fig:image2}
 \end{subfigure}
 \hfill
 \begin{subfigure}[t]{0.32\linewidth}
 \includegraphics[width=\textwidth]{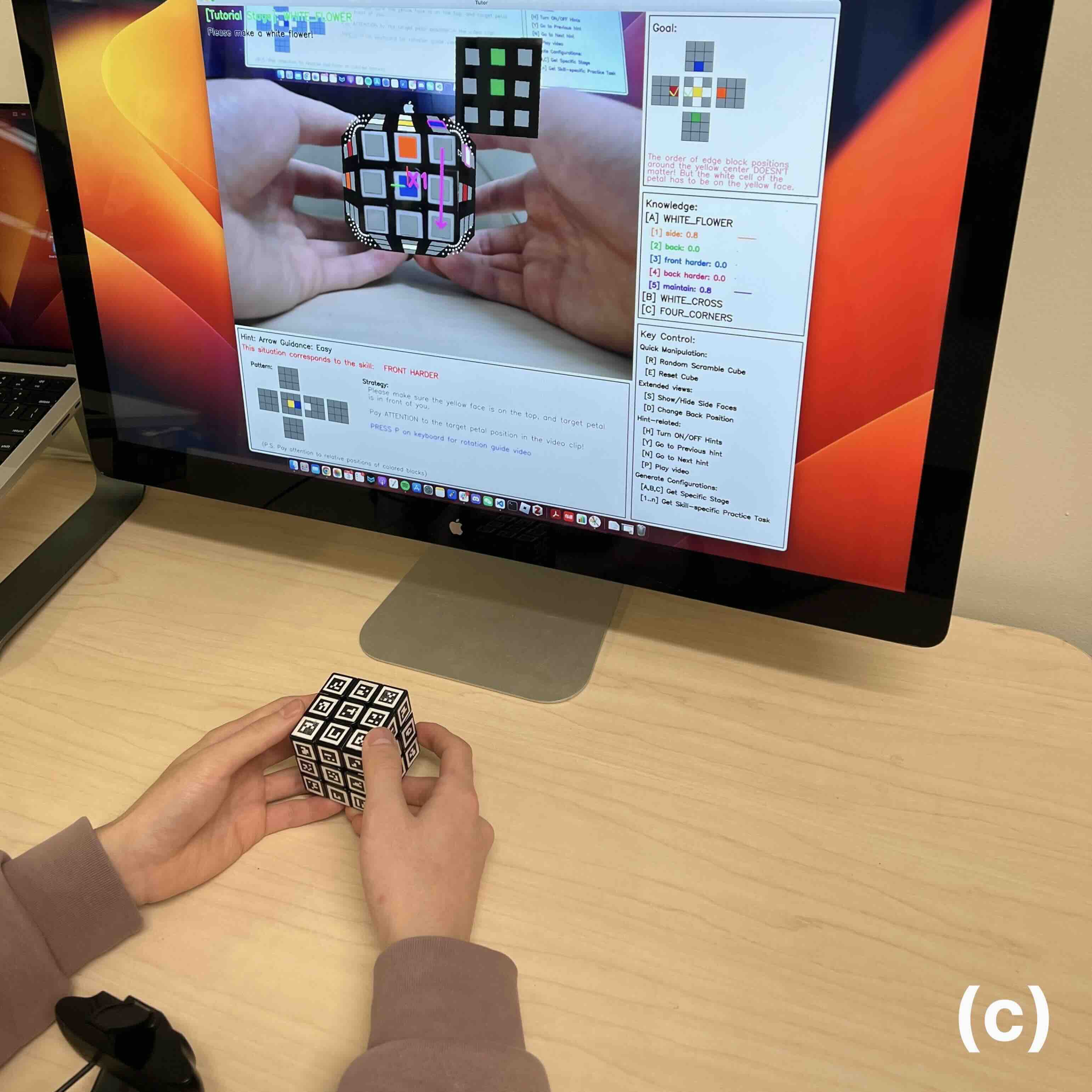}

 \label{fig:image3}
 \end{subfigure}
 
 \caption{Three conditions in the evaluation study. (a) Condition 1: Baseline 1 where users watch video tutorials and use a regular Rubik's Cube. (b) Condition 2: Baseline 2 where users watch video tutorials and use an AR-rendered cube. (c) Condition 3: Rubikon.}
\Description{Pictures of users learning to solve the Rubik's Cube in three ways labeled (a), (b), and (c). In Picture (a), the user holds a regular cube and watches the video tutorial on the display. In Picture (b), the user holds a Rubik's Cube with ArUco markers in front of a camera and watches the video on the display. In Picture (c), the user holds the same Rubik's Cube as the one in Picture (b), and they follow the Rubikon system on the display.}
\label{fig:conditions_setup}
\end{figure}

To evaluate Rubikon, we designed a between-subjects experiment in which participants learned to solve a Rubik's Cube in one of three conditions, as shown in Figure~\ref{fig:conditions_setup}. We adopted a business-as-usual baseline (Baseline 1) in which participants learned to solve the Rubik's Cube by watching online video tutorials, while manipulating a regular cube. In Baseline 2, participants manipulated an AR-rendered cube and watched the same online video tutorials as in Baseline 1. This allows us to investigate whether AR renderings would add excessive cognitive load to the learning process since the user's views and hand operations are separated.

\subsection{Participants}
We recruited 36 participants through mailing lists at a large public university in the United States. We included a screening survey in the original recruitment email to select participants who were novices on solving the Rubik's Cube. The survey contains several questions on solving a 3-by-3 Rubik's Cube. From a pool of 184 survey respondents who met the inclusion criteria of being a novice on the task, we sent email invitations following their response time to the screening survey. In the end, 36 people participated in the study, with 15 female, 20 male, 1 undisclosed, and an average age of 22, ranging from 18 to 33. 

\subsection{Study Procedure}
The study sessions lasted up to 2 hours in person. We administered a pre-test at the start of the session. Participants were presented with a goal image of a red cross and were asked to achieve this goal using a regular Rubik's Cube within 2.5 minutes. The time limit was set based on our pilot studies. Participants were scored based on their performance. Specifically, a correct square in place counts as one point, making the total score of the task 8 points. The pre-test served as a proxy of the participants' prior knowledge of solving the Rubik's Cube, which was also used to filter out those who were already experts. 

Participants were randomly assigned to one of the three conditions and were given 45 minutes to learn to solve the first layer of a Rubik's Cube. In the two baseline conditions, users could watch the tutorial video freely. We selected a popular online video tutorial \cite{youtubeSolveRubiks} with over 570K views that clearly breaks down the task into three stages. We further added captions to this online video tutorial to label the three stages and the 11 knowledge components as specified in the task model of Rubikon. Following the 45-min practice, participants answered the NASA TLX questionnaire \cite{NASATLX} to assess the cognitive load required by the task. We chose NASA-TLX because it is a widely used, validated tool for assessing cognitive workload in a variety of task contexts~\cite{yang2025videomix,nam2024using}. Participants then completed a post-test. We performed a short interview at the end to probe into the participants' learning experiences. For participants in the Rubikon condition, we asked their thoughts on the AR setup, the adaptive task generation functionality, and the perceived accuracy of the model tracing and knowledge tracing algorithms.

With the random assignment to conditions, there were 12 participants in each of the three conditions. The study was approved by our institution's IRB. All participants were compensated with a \$30 gift card.

\subsection{Learning Outcome Measure}
We administered a post-test after the practice asking participants to solve the first layer of a Rubik's Cube. We consider this to be an authentic post-test that reflects the tutoring objectives. We broke down the post-test into three tasks, aligned with the three learning objectives, namely solving White Flower, White Cross, and Four Corners. For each task, we configured the Rubik's Cube so that all participants went through the exact same post-test. Participants had 2.5 minutes to work on each task independently. We used a similar scoring mechanism as in the pre-test, i.e., when one square is back in place, it counts as one point, with a total score of 16 points. We also recorded the task completion time on the post-test. 

\subsection{Learning Process Measure}
\begin{figure}
    \centering
    \includegraphics[width=\linewidth]{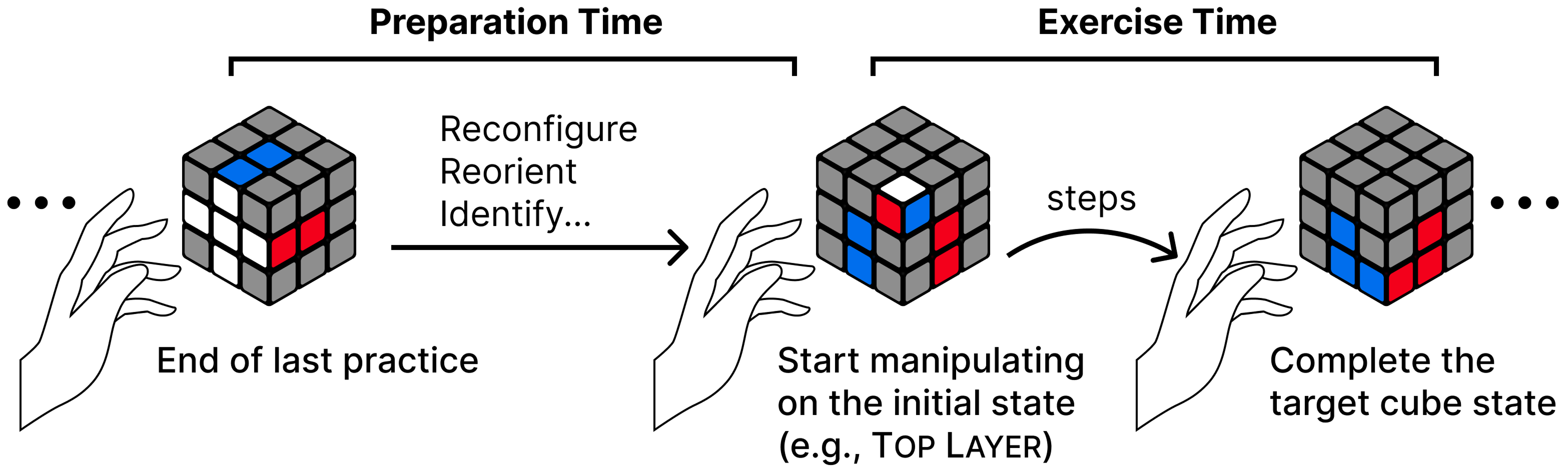}
    \caption{The learning processes of participants are divided into segments, each representing an exercise opportunity targeting a knowledge component defined by the starting and ending cube states.}
    \Description{An illustration showing the stages of a learning segment. It begins with a hand holding a cube at the end of a previous practice. During Preparation Time, the learner reconfigures, reorients, or identifies the cube state. During Exercise Time, the learner manipulates the cube starting from an initial state, such as solving the top layer, and continues until reaching a target cube state. The diagram is divided into Preparation Time and Exercise Time, with arrows showing progression between each step.}
    \label{fig:learning_process}
    \vspace{-1.5em}
\end{figure}

To understand how learners learned during the 45-min practice time, we defined a process measure ``exercise opportunity'' (see Figure~\ref{fig:learning_process}). Exercise opportunities are associated with each knowledge component. To label an exercise opportunity for a knowledge component, we need to identify the start and ending time of the exercise period. As introduced earlier, each knowledge component has a unique initial configuration and a target state. During practice, when the learner's cube matches the initial configuration of a knowledge component and the learner starts to change the position of the target block, it marks the start time for this exercise opportunity. When the target block is back in place, it marks the ending time. During this time interval, no other knowledge components would be exercised. This allows us to derive \textit{(i)} the number of knowledge components exercised during practice, and \textit{(ii)} the time duration for each exercise opportunity. 

During the practice time, a camera was positioned behind the users to record the study session, as shown in Figure \ref{fig:conditions_setup}. In the two baseline conditions, two researchers watched the video and manually labeled the exercise opportunities. In Rubikon, the system detects and logs the start and end of each exercise opportunity automatically. 

\subsection{Interview Analysis}
The interview recordings were transcribed and analyzed using affinity diagramming \cite{moggridge2007designing}. Two researchers interpreted the transcripts, iteratively grouped the interpretation notes, and identified emerging themes from the data. 

\begin{figure*}[t!]
 \centering
 \begin{subfigure}[h]{0.245\textwidth}
 \centering
 \includegraphics[width=1.0\textwidth]{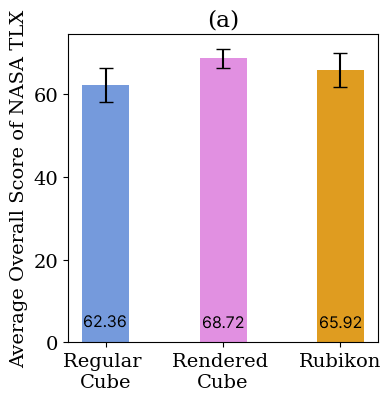}
 \end{subfigure}
 \hfill
 \begin{subfigure}[h]{0.245\textwidth}
 \centering
 \includegraphics[width=1.0\textwidth]{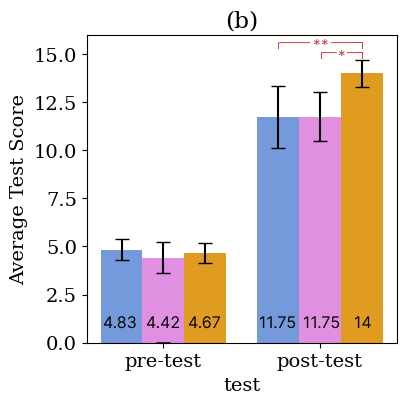}
 \end{subfigure}
 \hfill
 \begin{subfigure}[h]{0.245\textwidth}
 \centering
 \includegraphics[width=1.0\textwidth]{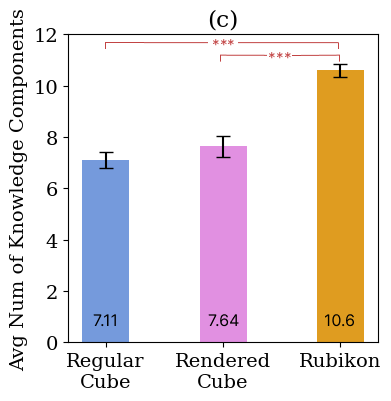}
 \end{subfigure}
 \hfill
 \begin{subfigure}[h]{0.245\textwidth}
 \centering
 \includegraphics[width=1.0\textwidth]{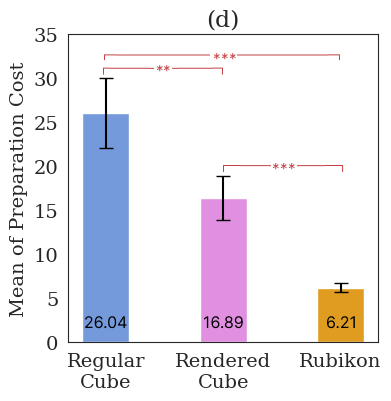}
 \end{subfigure}
 \vspace{-1pc}
 \caption{Evaluation Study Results. (a) Learners in the three conditions reported similar levels of NASA TLX overall score after the 45-min practice period. (b) Learners in Rubikon had significantly higher learning gains compared to learners in the two baselines. (c) 
 Learners in Rubikon exercised significantly more knowledge components during the 45-min practice time.
(d) Learners in Rubikon had significantly lower ``preparation cost'' --- the time required for cube reconfiguration to exercise a knowledge component.
*** = $p$ < 0.001, ** = $p$ < 0.01, and * = $p$ < 0.05.}
 \Description{This figure contains 4 plots. (a) is a bar plot of NASA Task Load Index overall scores for three conditions in phase 2. Each bar has a standard error bar. The average overall scores are shown at the bottom of three bars in the plots. Regular Cube Condition 1 has an average score of 62.36, Rendered Cube Condition 2 has an average score of 68.72, and Rubikon Condition 3 has average score 65.92. (b) Bar charts of the average scores of three conditions in the post-test and pre-test. Each bar has a standard error bar. In the pre-test, Regular Cube condition has average of 4.83, Rendered Cube condition has average of 4.42, and Rubikon has the average of 4.67. In the post-test, Regular Cube condition has average 11.75, Rendered Cube condition has average 11.75, and Rubikon condition has average 14. In post-test, there is a significance line connecting Regular Cube and Rubikon. It has two stars to significance level. There is a significance line connecting Rendered Cube and Rubikon with one star. (c) is a bar plot of average number of knowledge components exercised by users across three conditions. Regular Cube condition has average of 7.11, Rendered Cube condition has average of 7.64, and Rubikon condition has average of 10.6. There is a significance line connecting Regular Cube and Rubikon with three stars. There is a significance line connecting Rendered Cube and Rubikon with three stars. (d) is a bar plot of the average of preparation cost for users across three conditions. Regular Cube condition has average of 26.04, Rendered Cube condition has average of 16.89, and Rubikon condition has average of 6.21. There is a significance line connecting Regular Cube and Rubikon with three stars. There is a significance line connecting Rendered Cube and Rubikon with three stars. There is a significance line connecting Regular Cube and Rendered Cube with two stars.}
 \label{fig:4_figures}
\end{figure*}

\section{Results}
We report our results in response to each research question. Before we ran any analysis, we did a randomization check to make sure that there was no systematic difference in the participants' prior knowledge across the three conditions. A one-way ANOVA test on the pre-test scores suggested that there was no significant difference across the three conditions ($F(2,33) = 0.11, p = 0.90$).

\subsection{The AR Setup Did Not Add To Learners' Cognitive Load}
In order to understand whether the AR-rendered cube posed extra cognitive load on users, we ran an ANOVA test on the NASA TLX scores across the three conditions. Results showed that there was no significant difference on the cognitive load users reported across conditions ($F(2,33) = 0.79, p = 0.46$), as shown in Figure~\ref{fig:4_figures}a.

\subsection{Rubikon Led to Higher Learning Gains}\label{RubikonHighLearningGain}
In order to investigate the learning benefit of Rubikon, we built a linear regression model, with the post-test score as the dependent variable, and the condition as the fixed effect. We also included the learners' pre-test score, and the time they took on completing the post-test as covariates. We found that users in Rubikon had significantly higher learning gains compared to the two baselines, as shown in Figure~\ref{fig:4_figures}b. With the pre-test score and post-test task completion time controlled, the regression model suggested that, when a user was assigned to the Rendered Cube condition compared to the Rubikon condition, they on average achieved 2.9 points lower out of 16 points in the post-test ($t$ = 2.43, $p$ = 0.02); when a user was assigned to the Regular Cube condition, they on average achieved 3.8 points lower in the post-test ($t$ = 3.25, $p$ = 0.002). This was a considerable effect size as learners achieved 3-4 points higher out of 16 points (25\% of the whole task) through Rubikon in 45 minutes of the learning time. To explain what 25\% of the entire task might look like, it could mean being able to solve all four corners instead of just two, or being able to complete the full white cross compared to being unable to move any pieces of the cross back into place.

\subsection{Learners in Rubikon Had a More Comprehensive and Balanced Coverage of Knowledge Components During Practice}
We further analyzed learners' learning processes to understand the reasons why users in the Rubikon condition learned more than the two baselines. 

\subsubsection{Learners exercised more knowledge components in Rubikon}\label{moreKC}
We examined the average number of knowledge components exercised by users in each condition. Our hypothesis was that since Rubikon generated new problem-solving activities based on learners' mastery, learners might get to exercise more knowledge components during the practice session. We ran a one-way ANOVA test and found there was a significant difference across the three conditions on the number of knowledge components exercised ($F$ = 29.52, $p$ < 0.0001). 
A pairwise post-hoc Tukey's HSD test indicated that Rubikon learners exercised significantly more knowledge components compared to the two baselines ($p$ < 0.0001 for both), as shown in Figure \ref{fig:4_figures}c. On average, learners in Rubikon exercised 3 more knowledge components in comparison to the baselines. This suggested that learners in Rubikon practiced in more diverse situations than what they would have seen from a video tutorial. 

\subsubsection{Learners in Rubikon distributed their time more evenly among knowledge components}\label{moreOften}
\begin{figure}[t!]
 \includegraphics[width=0.9\linewidth]{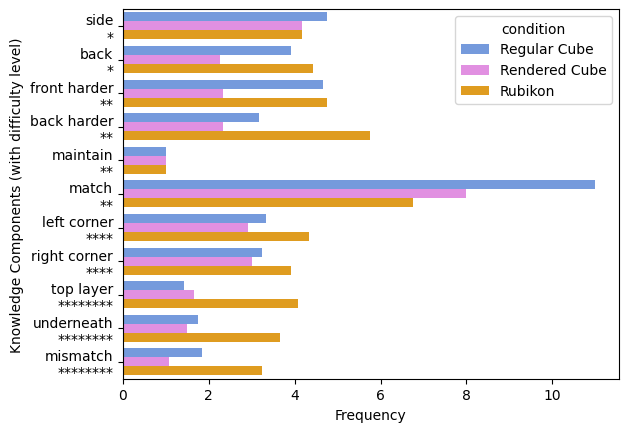}
 \vspace{-1pc}
 \caption{The frequency users exercised each knowledge component across the three conditions. * in the y-axis indicates the difficulty level of each knowledge component (see definition in Section \ref{learningObjectives}). Rubikon learners had a more balanced coverage in their practice and a bigger edge in comparison with the baselines on the difficult knowledge components.}
 \Description{This figure is a bar plot of the frequency that users exercised each knowledge component across three conditions. The knowledge components on the y axis have stars underneath them, and the number of stars indicates their difficulty level. The x axis indicates the frequency of each knowledge components. First five knowledge components belong to White Flower stage. One knowledge component after the first five belongs to White Cross. The last five knowledge components belong to Four Corners stage. The knowledge components are also ordered from top to bottom with ascending difficulty level. The bars for Regular Cube and Rendered Cube condition have higher frequency values on easy knowledge components. The bars for Rubikon condition have more evenly distributed frequency values. One knowledge component ``match'' has the highest frequency values in all three conditions compared to other knowledge components.}
 \label{fig:skill_distribution}
\end{figure}

Figure~\ref{fig:skill_distribution} shows the average frequency participants exercised each knowledge component across the three conditions. The number of asterisks indicated the difficulty level of each knowledge component (see definition in Section \ref{learningObjectives}). A Shapiro-Wilk test showed that the distribution of users' exercise opportunities in the Rubikon condition followed a normal distribution ($p$ = 0.35) whereas the two baseline conditions were skewed towards the easier knowledge components ($p$ < 0.001 for both tests).

Notably, learners in Rubikon exercised the difficult knowledge components more than the two baselines. We observed that learners in the two baseline conditions did not want to reconfigure the Rubik's Cube manually to exercise a specific knowledge component because reconfiguration might mess up a step they had completed. Moreover, learners also did not know how to reconfigure the Rubik's Cube manually to a specific state. In these cases, learners could only watch the video tutorial without hands-on practice. In contrast, in Rubikon, users got prompted to exercise every knowledge component and were given repeated practice opportunities when they had not demonstrated independent success on them.

\subsubsection{Learners in Rubikon spent less time on preparation and configuration before each exercise opportunity}
When learning to solve the Rubik's Cube with video tutorials, learners needed to align their cube with the status of the cube in the video. The mapping of the two representations took skill, effort and time. We further labeled ``exercise time'' and ``preparation time'' (see Figure~\ref{fig:learning_process}). ``Exercise time'' is the duration of an exercise opportunity, and ``preparation time'' is the interval between the previous and current exercise opportunities.
Since the time spent configuring the Rubik's Cube to a specific state constitutes ``preparation'' rather than ``learning'', we introduce the metric ``preparation cost'', defined as the ratio of ``preparation time'' to ``exercise time.'' This metric allows us to evaluate whether Rubikon improves learning efficiency by reducing the time needed to configure the cube. When the ``preparation cost'' is lower, it means the learners need lower preparation effort to engage in meaningful practice. We ran a one-way ANOVA test and found there was a significant difference across the three conditions on the ``preparation cost'' ($F$ = 28.87, $p$ < 0.001). A pairwise post-hoc Tukey's HSD test suggested that Rubikon users had significantly lower ``preparation cost'' in comparison with the two baselines ($p$ < 0.0001 for Regular Cube condition, $p$ = 0.0003 for Rendered Cube condition), as shown in Figure~\ref{fig:4_figures}d. The AR setup enabled the system to automatically reconfigure the rendered Rubik's Cube so that the learners did not need to waste time reconfiguring the cube to a specific state.

\subsection{Subjective User Experiences}
We summarized the themes from the affinity diagram of the interview transcripts. 

\subsubsection{The AR setup incurred higher cognitive load.}
Even though we did not see a difference in the cognitive load users reported through the NASA TLX questionnaire, in the interviews, participants did mention that the AR setup introduced higher cognitive demand. Some participants shared that the AR setup constrained the way they manipulated the cube, which resulted in higher physical and mental efforts. On the other hand, multiple participants shared that the cognitive demand decreased as they became familiar with the system. For example, P10 commented, \textit{``Rubikon was new so it took some time before I could get used to it and use it as well as the normal Rubik's Cube.''}

\subsubsection{Extended views provided useful information.}
Multiple participants shared that they were able to use the extended views to reconstruct the same 3D space as when using a regular cube, specifically with the visual aids Rubikon provided. \textit{``I think I got the point where the virtual cube has the same spatial awareness as the physical cube. (P14)''} Many participants also consistently mentioned that the extended views provided just enough information that was easily understandable, \textit{``Helpful to present more information, not too overwhelming, straightforward. (P7)''} With the extended views, participants could quickly locate the squares they were looking for without the need to physically manipulate the cube to another side. On the other hand, some participants shared concerns about the usability of the extended views. P22 and P35 mentioned that the squares in the extended views were too small, and P29 said that \textit{``it was hard for me to visualize the blocks on the back side.''}

\subsubsection{Learners appreciated the automatic configuration of the Rubik's Cube in Rubikon.}
In the baseline conditions, users had to match the configuration of the Rubik's Cube in their hands with that in the tutorial video, which required time, effort, and skill. Many participants found this process frustrating due to mismatches between the cube in their hand and the one in video. \textit{``When mismatches occur, I don't know how I can get there. (P27)''} They struggled to progress smoothly through the tutorial, as it presented the solving process linearly and chronologically, resulting in either wasted time or incomplete understanding.

In contrast, Rubikon users did not face such issues. They appreciated Rubikon's automatic generation of problems targeting each knowledge component (P1, P25) and valued the additional practice opportunities it provided, leading to a more efficient learning experience (P34).

\subsubsection{Knowledge tracing provided valuable practice opportunities to learners.}
Participants found the knowledge tracing functionalities of Rubikon to be valuable. Several participants pointed out that the knowledge tracing algorithm was able to track the knowledge components they did not master well. \textit{``I felt like I did struggle with those low scores skills. (P4)''} Participants (P13, P34) also found the skillometer functionality to be useful. P34 mentioned, \textit{``I think it was useful to know which skills I still need practice with so I could pay more attention during the tutorial.''}

\section{Discussion and Future Work}

\subsection{Generalizability of Rubikon}
The study revealed that utilizing the proposed AR setup to generate new Rubik's Cube configurations led to more comprehensive coverage of knowledge components during practice, resulting in enhanced learning. For example, the system generates configurations of the Rubik's Cube where a target square is placed at varying positions so that learners could practice mapping the square back in more diverse situations than what they would have seen from a video tutorial. Learners experienced a reduced ``preparation cost'' for engaging in meaningful practice, meaning that they did not need to spend time aligning their physical Rubik's Cube to a virtual cube in the video tutorial. We want to emphasize that Rubikon is a Rubik's Cube-specific solution, and the AR setup and the behavior tracking techniques may not generalize beyond the task of learning to solve a Rubik's Cube. On the other hand, this work investigates the feasibility and pedagogical value of automatically generating targeted and personalized practice opportunities for physical task learning while retaining the haptic experience. We consider the concept of physical task reconfiguration applicable in other mixed reality-based physical task tutoring and guidance systems. We will discuss future work directions next. 

\subsection{Beneficial Situations to Repeatedly Generate Configurations of Physical Objects}
Based on the case of solving a Rubik's Cube, we lay out some properties of a 3D physical task where the generation of new configurations of physical objects could be helpful.

\begin{enumerate}
 \item \textbf{The spatial layout is important for learning.} For tasks where the spatial layout is important, e.g., solving a Rubik's Cube, playing chess, etc., generating new configurations that represent different spatial layouts could help learning. 
 \item \textbf{Learners want to jump to intermediate states in a procedure to practice.} When it is critical for learners to be able to jump to any intermediate state of a procedure to practice, the capability to reset the physical task to any state would be beneficial. For example, in the case of learning drawing, learners can work on half-completed drawings (rendered digitally or by AR) to deliberately practice a skill instead of having to start from scratch every time. 
 \item \textbf{One solution path does not necessarily contain all the knowledge components encompassed in the task.} When the task is complex and has multiple solution paths, learners may not easily get exposed to all the required knowledge components if they just practice on their own. For example, when someone is learning to use a physical interface that has diverse paths, using AR to project new configurations on a simulated physical interface could help people more easily exhaust the possible scenarios.
 \item \textbf{The task requires complex or expensive configurations of physical objects.} When the task requires complex or expensive configurations of physical objects, AR-enabled reconfiguration could save both time and resources.
\end{enumerate}

\subsection{Implications for Physical Task Tutoring}
In this work, we present evidence on the benefit of automatically configuring a Rubik's Cube in order for learners to more efficiently practice and learn to solve it. In particular, a main reason Rubikon is more helpful than traditional tutorials is that learners can easily align the state of the physical object they are manipulating with the one in the tutorial. We observed struggles and frustrations from the users when they could not match the Rubik's Cube in their hands with the one in the video tutorial. When using AR to present different configurations of the physical objects required in the physical tasks, it also offers an opportunity to provide extra information for the users which is inaccessible without AR. In Rubikon, we showed users the hidden views of a Rubik's Cube and used simple visualizations to make them more comprehensible. We believe this notion can be extended to other domains, even though they may require a different AR setup from Rubikon (e.g,. ArUco markers) for tracking and display that fits the nature of the task.

Although there has been a plethora of work in both the field of intelligent tutoring systems \cite{abueloun2017mathematics, al2017design, yamaguchi_video-annotated_2020}, and AR-based tutorial systems for physical tasks \cite{thoravi2019loki, faridan2023chameleoncontrol, monteiro2023teachable, anderson2013youmove, gupta_duplotrack_2012}, there has been limited work integrating both, and limited empirical studies examining the learning benefit of such tutors. One example is Origami Sensei \cite{chen2023origami}, which uses computer vision and real-time projection to guide learners through structured, planar origami tasks. Although it showcases the potential of AI-augmented MR systems, its scope is restricted to relatively linear tasks with fixed action sequences. We carefully designed Rubikon following the design principles of cognitive tutors, which has witnessed huge success in disciplines such as math \cite{olsen2014using, koedinger1997intelligent, koedinger2007exploring, rau2009intelligent} and programming \cite{rivers2017data, price2017isnap} over the past decade. In our study, the knowledge and model tracing algorithms were applauded by the participants for offering valuable feedback and directing them to useful practice opportunities. We encourage more work to deeply integrate MR technologies with theory-driven instructional design to produce effective tutoring systems for 3D physical task learning. 

\subsection{Limitations}
Our work has several limitations. Firstly, despite the comparable NASA TLX score, Rubikon's on-screen display for rendering AR objects may still increase cognitive load, as some users reported. We plan to explore head-mounted displays in the future, expecting improved coordination between learners' views and hand operations, potentially reducing cognitive load further. Secondly, our evaluation study compares Rubikon with conventional approaches to physical task learning. However, for the Rubik's Cube, there are web-based platforms allowing practice through GUI-based interactions, as well as smart cubes that track the cube's state (as mentioned in Section \ref{sec:rubikscubelearning}). We did not include a baseline where learners use such platforms, though the benefits of Rubikon's AR-enabled task reconfiguration remain applicable. Thirdly, due to time constraints in the user study, we evaluated Rubikon's performance only on solving the first layer of the Rubik's Cube. The current system only specifies the task model for the first layer. However, we believe our findings are generalizable to learning other parts of the Rubik's Cube. Future studies could adopt longer instruction and assessment time spans to examine the effects of intelligent tutoring on long-term skill retention.

\section{Conclusion}
We presented Rubikon, an intelligent tutoring system for learning to solve the Rubik's Cube. The foundational design of Rubikon is an AR setup, where learners manipulate a physical cube while seeing an AR-rendered cube. This allows Rubikon to automatically generate configurations of the Rubik's Cube to target learners' weaknesses and help them exercise diverse knowledge components. In the evaluation study, we found that the users learned more in Rubikon in comparison to a business-as-usual baseline, where they watched video tutorials while manipulating a physical cube. We found that learners in Rubikon exercised more knowledge components and had lower ``preparation cost'' to engage in meaningful practice. Beyond solving a Rubik's Cube, we identified situations when repeatedly generating configurations of physical objects could help with the learning of physical tasks. This includes tasks that have an emphasis on spatial layout, require learners to easily jump to intermediate states, have multiple diverse solution paths, and require complex or expensive configuration of physical objects. 

\balance
\bibliography{citation}
\bibliographystyle{ACM-Reference-Format}

\end{document}